\documentclass[%
 reprint,
nofootinbib,
 amsmath,amssymb,
 aps,
]{revtex4-2}

\usepackage{graphics}
\usepackage{amsmath}
\graphicspath{{Figures/},{Logos/}}
\usepackage{graphicx,color}

\usepackage{bm}
\usepackage{amsfonts}
\usepackage{amsmath}
\usepackage{amssymb}
\usepackage{float}
\usepackage{hyperref}
\usepackage{subfigure}
\usepackage{slashed}

\usepackage{multirow}

\newcommand{\be}{\begin{equation}}
\newcommand{\ee}{\end{equation}}
\newcommand{\bea}{\begin{eqnarray}}
\newcommand{\eea}{\end{eqnarray}}

\newcommand{\tev}{\rm{TeV}}
\newcommand{\gev}{\rm{GeV}}
\newcommand{\mev}{\rm{MeV}}
\newcommand{\kev}{\rm{keV}}

\newcommand{\calL}{{\cal L}}

\begin{document}

\title{The neutron decay anomaly, neutron stars and dark matter}




\author{Mar Bastero-Gil} \email{mbg@ugr.es} \affiliation{Departamento
  de F\'{\i}sica Te\'orica y del Cosmos, Universidad de Granada,
  Granada-18071, Spain}
\affiliation{Laboratoire de Physique Subatomique et de Cosmologie,Universit\'{e} Grenoble-Alpes, CNRS/IN2P3, 38000 Grenoble, France}

\author{Teresa Huertas-Rold\'an} \email{thuertas@iac.es} \affiliation{
Instituto de Astrof\'{\i}sica de Canarias, C/V\'{\i}a Láctea 1, E-38205, La Laguna, Spain} \affiliation{Departamento de Astrofísica, Universidad de La Laguna, E-38206 La Laguna, Spain}

\author{Daniel Santos} \email{daniel.santos@lpsc.in2p3.fr} \affiliation{Laboratoire de Physique Subatomique et de Cosmologie,Universit\'{e} Grenoble-Alpes, CNRS/IN2P3, 38000 Grenoble, France}

\begin{abstract}
  The discrepancies in different measurements of the lifetime of isolated neutrons could be resolved by considering an extra neutron decay channel into dark matter, with a branching ratio of the order of $O(1$\%). Although the decay channel into a dark fermion $\chi$ plus visible matter has  already been experimentally excluded, a dark decay with either a scalar or dark photon in the final state remains still a possibility. In particular, a model with a fermion mass $m_\chi\approx 1$ GeV and a scalar $m_\phi \approx O(\mev)$ could provide not only the required branching ratio to explain the anomaly but also a good dark matter (DM) candidate with the right thermal abundance today. Although the interaction DM-neutron will affect the formation of neutron stars, the combined effect of the dark matter self-interactions mediated by the light scalar and an effective repulsive interaction with the neutrons induced by the scalar-Higgs coupling would allow heavy enough neutron stars. Combining the constraints from neutron lifetime, dark matter abundance, neutron stars, Higgs physics, and Big Bang Nucleosynthesis, we can restrict the light scalar mass to be in the range $2 m_e < m_\phi < 2 m_e + 0.0375$ MeV.

\end{abstract}

\maketitle

\section{Introduction}

The Standard Model (SM) of particle physics provides a successful description of the fundamental particles and their interactions. Nevertheless, we know that this is still an incomplete picture, as it does not explain for example the nature of Dark Matter (DM), the origin of the asymmetry matter-antimatter, or that of light neutrino masses. Besides, there are still open questions regarding some particle measurements, like the lifetime of the neutron, one of the building blocks of baryonic matter. The isolated neutron mainly decays via $\beta$ processes into protons, electrons and anti-neutrinos:
\begin{equation} \label{eq:beta_decay}
n \rightarrow p + e^- + \bar{\nu}_e \,.
\end{equation}
Two different methods have been used to measure this lifetime. The bottle or storage method \cite{Mampe1993,Serebrov2005,Serebrov2008,Pichlmaier2010,Steyerl2012,Arzumanov2015} counts the number of neutrons that have not decayed, with an average lifetime value $\tau_n^{\text{bottle}} = 878.4 \pm 0.6 \,$s  \cite{ParticleDataGroup2022}. On the other hand, the beam method \cite{Byrne1996,Yue2013,Sumi:2021svn} counts protons from $\beta$ decay, with an average $\tau_n^{\text{beam}} = 888.0 \pm 2.0 \,$s  \cite{Sumi:2021svn}.
The results are highly dependent on the method \cite{Rajan:2018hii}, and this discrepancy of $9.6$ s or $4.0\,\sigma$ is known as the neutron decay anomaly. Although it is not excluded that the difference is due to (yet unknown) systematic errors \cite{Giacosa2020}, it might also be due to a new decay channel unaccounted for in beam experiments. For example, as proposed in Ref. \cite{Fornal2018}, the discrepancy could be explained with a new neutron decay channel into a DM fermion $\chi$:
\be
n \rightarrow \chi + {\rm visible} ({\rm invisible}) \,,
\ee
where ``visible'' can be either a photon or a pair of $e^-e^+$, and ``invisible'' either a dark photon ($A^\prime$) or a dark scalar ($\phi$). Explaining the difference in the neutron lifetime would require a branching ratio ${\rm Br}(n \rightarrow \chi \,+\,{\rm anything}) \approx 1\%$. The ``visible'' channel with a photon was soon experimentally excluded \cite{Tang:2018eln}, and the one with a pair of $e^-e^+$ constrained for kinetic energies $37.5 \,\kev \lesssim E_{e^+e^-} \lesssim 644 \,\kev$ (95 \% CL) \cite{UCNA:2018hup,Klopf:2019afh}. For the invisible dark decay channel, the proposal was questioned related to  Neutron Stars (NS) physics \cite{tovns1, tovns2, tovns3}. The mixing between neutrons and dark fermions leads to the partial conversion of neutrons to DM inside the star, softening the equation of state (EoS) at high densities, and lowering the possible maximal mass of the NS below the largest observed mass  $M_{NS}\sim 2 M_{\odot}$ \cite{Riley2021}. However, as it was already noticed in \cite{tovns1}, DM repulsive self-interactions could avoid the problem. For example in \cite{Cline2018} the dark sector was assumed to be charged under an extra $U(1)$ symmetry, and extended including a dark photon $A^\prime$. The dark repulsive Coulomb-like interaction was enough to compensate for the softening of the EoS and to recover large enough NS masses; however different cosmological, astrophysical and particle physics constraints preclude the possibility of the dark fermion accounting for the whole DM abundance in this model. Another possibility was explored in \cite{Grinstein2019} where they considered the invisible decay channel with a singlet dark scalar. Through the mixing of the dark scalar with the Higgs, an effective repulsive DM-neutron interaction can be generated, which again would disfavour the production of DM inside the NS; this model could also account for the right DM relic abundance. Another simpler possibility to recover massive enough NS and the dark matter relic abundance was proposed in \cite{Strumia:2021ybk, Husain:2022brl}, which considered the decay into 3 dark fermions. 

In this work we focus on the model presented in \cite{Grinstein2019} with a light scalar mediator $\phi$, and their explicit particle model realisation including the coupling to the Higgs sector. They worked with $m_\phi \approx O(0.1)$ eV, which requires extending further the model to ensure the scalar decays before Big Bang Nucleosynthesis (BBN); otherwise it would contribute to the effective number of relativistic degrees of freedom at the time of BBN, which is severely constrained by cosmological observations \cite{Pitrou:2018cgg, Planck2020}. On the other hand, a scalar mediator with a mass $m_\phi > 2 m_e$ could decay into a pair $e^-e^+$,  avoiding BBN constraints. However, this will indeed reintroduce the neutron decay channel with a pair $e^-e^+$ in the final state 
\be
n \rightarrow \chi + \phi^* \rightarrow \chi + e^- + e^+ \,.
\ee
Although this has been constrained to have a branching ratio less than 1\% for $e^-e^+$  kinetic energies in the range $37.5 \,\kev \lesssim E_{e^+e^-} \lesssim 644 \,\kev$ at 95 \% CL, still there is an unexplored region below and above this range, and we could  have an invisible channel either with a dark light scalar with a mass $2 m_e < m_\phi \lesssim 2m_e + 37.5$ keV, or $m_\phi \gtrsim 2m_e + 644$ keV. We stress that from the point of view of the scalar sector, its decay into SM fermions, besides being the simpler solution to the potential problems at BBN, may also open new venues of searching for possible experimental signatures either in neutron/baryon physics or Higgs physics. 

We will thus revise the different constraints on model parameters to explain the neutron decay anomaly with a DM fermion $m_\chi \approx m_n$ and a singlet scalar mediator $m_\phi \approx O(\mev)$, such that (a) it is compatible with the observed DM abundance;  and  (b) it allows for a repulsive (Yukawa potential) interaction between the neutron and the DM fermion and therefore massive enough NS. In addition DM self-interactions mediated by a light scalar with  $m_{\phi}\sim O(\mev)$ might help to  solve the small scale structure problems \cite{Tulin2013a, Tulin2013,Kaplinghat2013, Kahlhoefer2017}. 
In section \ref{particlemodel} we will first review the microscopic model proposed in \cite{Grinstein2019} in order to set the notation and mass ranges. In section \ref{DMabundance} we compute the DM abundance from the standard freeze-out mechanism, and set the value of the coupling between the DM fermion and the light singlet scalar $g_\chi$. Constraints from Higgs physics are revised in section \ref{ScalarPotential}, and those from NS physics in section \ref{NSDM}. Finally in section \ref{Summary} we summarize all the constraints and discuss the parameter space available. 
  In Appendix A we comment on how the parameter space is modified when slightly reducing the value of $g_\chi$, which could be favoured by small-scale structure considerations for our Self-Interacting DM (SIDM) candidate. Finally in Appendix B we just quote the expressions for the dark neutron decay.


\section{Particle Physics Model} \label{particlemodel}

The particle physics model proposed in \cite{Grinstein2019} is a modification of the one presented in \cite{Fornal2018}, and it requires two extra scalars and  fermions: a heavy scalar $\Phi=(3,1)_{-1/3}$ with baryon number $B_\Phi$=-2/3, which mediated the neutron decay;  a light scalar singlet $\phi$ with $B_\phi=0$ coupled to the Higgs, which generates the repulsive interaction; the DM Dirac fermion $\chi$, and an extra Dirac fermion $\tilde \chi$ heavier than the neutron, with $B_\chi=B_{\tilde \chi}=1$. At the quark level, the interactions are given by:
\bea
  {\mathcal L} &&= \left(\lambda_q\,\epsilon^{ijk}\,\overline{u_{Li}^c}\,d_{Rj}\,\Phi_k + \lambda_{\chi}\,\Phi^{*i}\,\overline{\tilde{\chi}}\,d_{Ri} \right.\nonumber \\
&&  + \left.\lambda_l\,\overline{Q_{Ri}^c}\,l_L\,\Phi^{*i} + \lambda_Q\,\epsilon^{ijk}\,\overline{Q_{Ri}^c}\,Q_{Lj}\,\Phi_k + \text{h.c.}\right) \nonumber\\
&&- M_{\Phi}^2|\Phi|^2 - m_{\tilde{\chi}}\,\overline{\tilde{\chi}}\,\tilde{\chi}  - m_{\phi}^2|\phi|^2 - m_{\chi}\,\overline{\chi}\,\chi \nonumber \\
  &&  + \lambda_{\phi}\,\overline{\tilde{\chi}}\,\chi\,\phi + \mu\,H^{\dagger}H\,\phi + g_{\chi}\,\overline{\chi}\,\chi\,\phi + g_{\phi H}\,\phi^2H^{\dagger}H
  \,.
\label{Lparticles}
\eea
The trilinear interaction $\mu$ between the singlet and the Higgs gives rise to an effective $g_n \bar n n \phi$ vertex, with $g_n= \mu \sigma_{\pi n}/m_h^2$, where  $m_h$ is the Higgs mass and $\sigma_{\pi n}=\sum_q \langle n|m_q\bar{q}q|n\rangle \approx 370\,$MeV \cite{Finkbeiner2008,Alarcon:2011zs}, with the sum running over all quark flavours. This originates the repulsive (with $\mu < 0$) effective potential between the neutron and the DM thorough the exchange of the light scalar $\phi$:
\be
V= \frac{g_\chi |g_n|}{4 \pi} \frac{e^{-m_\phi r}}{r} \,. \label{Yukawapot}
\ee

Except for the masses of $\chi$ and $\phi$, we assume that other particles in the model are always heavy enough to avoid constraints (if any) from accelerator physics or cosmology.
In order to allow for  the neutron decay channel into $\phi + \chi$, but ensuring the stability of nuclei \cite{Pfutzner2018}, one must impose :
\be
  937.993 ~\mev< m_\chi + m_\phi < 939.565 ~ \mev \,, \label{masses1}
\ee
where the upper limit is the neutron mass and the lower limit avoids the decay of $^9Be$. One must also impose $m_{\tilde \chi} > 937.993$ MeV to avoid the decay of $^9Be$ through $n \rightarrow \tilde \chi \gamma$. Eq. \eqref{masses1} ensures the stability of the proton, and to prevent the possible decay of the dark fermion into protons, i.e, $\chi \rightarrow p^+ + e^- + \phi$, one needs:
\be
  | m_\chi - m_\phi | < 938.783 ~ \mev \,. \label{masses2}
\ee
Instead of working with a light singlet scalar $m_\phi \lesssim 2 m_e$, which would require further extending the model to ensure their decay before BBN, we will work with $m_\phi \approx O(\mev)$, excluding the values constrained by 
the limits on the induced decay $n \rightarrow \chi + e^-+e^+$ set in \cite{UCNA:2018hup,Klopf:2019afh}. The DM fermion mass is then $m_\chi \sim O(1\, \gev)$, within the range given in Eq. \eqref{masses1}. The condition in Eq. \eqref{masses2} is always fulfilled, and the fermion $\chi$ is stable and a good candidate for DM.  

About the couplings, neutron dark decay rate will depend on the combination of couplings $\lambda_\phi \lambda_q \lambda_\chi$ \cite{Fornal2018} (see Appendix A), and requiring the branching ratio to be less than 1\% only gives a mild constraint $10^{-5} \lesssim \lambda_\phi \lambda_q \lambda_\chi \lesssim  10^{-2}$ with $M_\phi \simeq m_{\tilde \chi} \simeq 1 \, \tev$. The couplings in the last line in \eqref{Lparticles} will be set in the following sections demanding the model to be consistent with the DM abundance, Higgs and NS physics.

\section{Dark matter abundance} \label{DMabundance}

The DM abundance can be derived from the standard freeze-out mechanism, integrating the Boltzmann equation \cite{Steigman2012}
\be 
\frac{dn_{\chi}}{dt} = -3Hn_{\chi} - \langle\sigma v \rangle\left[n_{\chi}^2 - (n_{\chi}^{eq})^2\right] \,, \label{dnchi}
\ee
where $n_{\chi}^{\text{eq}}$ is the number density in equilibrium, $H$ the Hubble parameter and $\langle\sigma v\rangle$ the thermal averaged cross section, usually parameterised as
\be \label{eq:cross-section_parametrisation}
\langle \sigma v\rangle = \frac{1}{m_{\chi}^2}\left(\alpha_s + \alpha_p \frac{T}{m_{\chi}}\right) \,, 
\ee
where $\alpha_s$, $\alpha_p$ are dimensionless parameters associated with s-, p-waves annihilation channels respectively.  For the model given in Eq. \eqref{Lparticles},  and assuming $m_{\tilde \chi} \gg m_\chi$ and/or $\lambda_\phi \ll 1 $, the DM annihilation cross-section into light scalars is given by \cite{Kouvaris2015}
\begin{equation} \label{eq:cross-section_annihil}
  \sigma_{\bar{\chi}\chi\rightarrow\phi\phi} v_{\text{rel}} \simeq \frac{3g_{\chi}^4}{32\pi m_{\chi}^2}\cdot \frac{T}{m_\chi} \,,  
\end{equation}
and therefore $\alpha_p= 3 g_\chi^4/(32\pi)$. The observational value of the DM density parameter $\Omega_{DM}$ \cite{Planck2020}, 
\be \label{eq:DM_abundance_Planck}
\Omega_{DM}h^2 = 0.120 \pm 0.001 \,, 
\ee
where $h$ is the reduced Hubble parameter, is reproduced with $\alpha_p = 1.78 \cdot 10^{-7} m_\chi^2$, which for $m_\chi \simeq 1$ GeV implies $g_\chi \simeq 0.05$. We notice that the freeze-out mechanism and Eq. \eqref{dnchi} assume implicitly that both the dark matter and the annihilation products are initially in thermal equilibrium. For the singlet scalar, this can be achieved before electroweak symmetry breaking through the scalar-Higgs interaction with a coupling value $g_{\phi H} \gtrsim 10^{-7}$ \cite{Enqvist2014}. The DM fermion is then kept in equilibrium with the singlet until it decouples at a temperature $T_D \approx m_\chi/20 \simeq 50$ MeV. 

\begin{figure}[t]
\centering
\includegraphics[width=8.5cm]{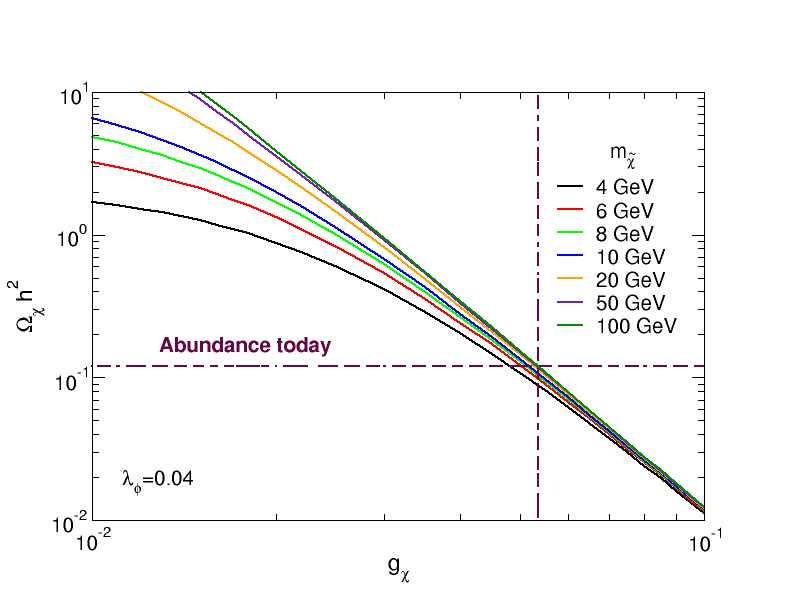} 
\includegraphics[width=8.5cm]{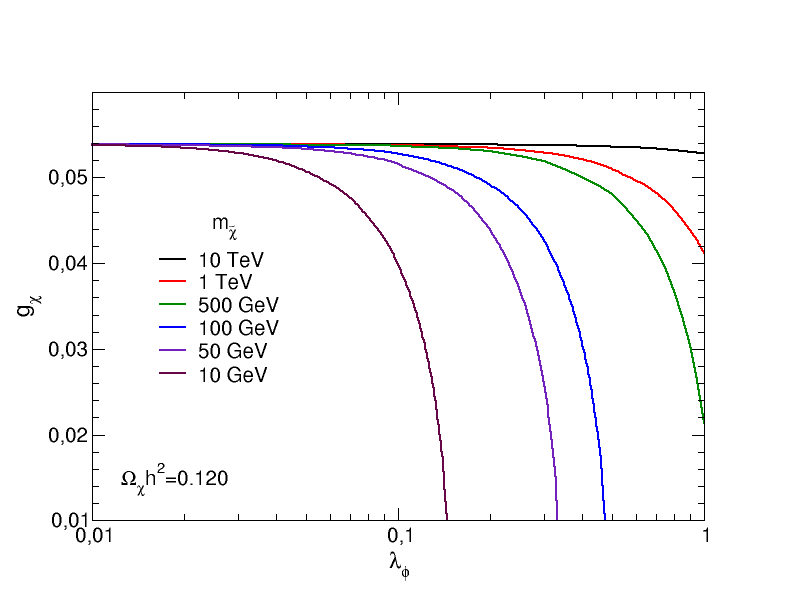}
\caption{\small Top panel: DM relic abundance obtained with micrOMEGAS varying $g_\chi$, for $\lambda_\phi =0.04$ and different values of $m_{\tilde \chi}$ as indicated in the plot; the vertical dashed line indicates the value $g_\chi \simeq 0.053$. Bottom panel: Relation between $g_\chi$-$\lambda_\phi$ to obtain the observed DM abundance today, for different values of $m_{\tilde \chi}$. See the text for other model parameters.}
\label{figMicromegas}
\end{figure}

  In order  to check the abundance dependence on other model parameters, we have also computed the relic abundance with micrOMEGAS \cite{Belanger2010, Alguero:2023zol}. The coloured scalar $\Phi$ will not play any role as far as it is heavy enough, so we have set $M_\phi=1$ TeV and have not included in the analyses the interactions given in the first two lines in the Lagrangian in Eq. \eqref{Lparticles}. We include the interaction terms depending on the couplings $g_\chi$, $\lambda_\phi$, $g_{\phi H}$, the trilinear coupling $\mu$, and the mass  of the heavier fermion $\tilde \chi$. Given that the decoupling temperature $T_D$ would be much smaller than the Higgs mass, we did not find as expected any dependence on the parameters of the singlet scalar-Higgs sector $\mu$ and $g_{\phi H}$, and to explore the dependence with the other parameters we have set $g_{\phi H}=5\times 10^{-3}$ and $\mu=40$ GeV. In the top panel of Fig. \eqref{figMicromegas} we have plotted the DM abundance obtained by varying $g_\chi$ for different values of $m_{\tilde \chi}$, and $\lambda_\phi=0.04$. For $m_{\tilde{\chi}} \gtrsim 50\,$GeV, the heavier fermion $\tilde{\chi}$ does not have any impact on the relic abundance calculation, and we obtain the observed abundance with $g_\chi \simeq 0.053$, same value than from our Boltzmann calculation. However, as the ratio $\lambda_\phi^2/m_{\tilde \chi}$ increases, processes mediated by the $\tilde \chi$ fermion may contribute to the total annhilitation cross-section, such that today's DM abundance can be recovered for smaller values of $g_\chi$, as seen in the bottom panel in Fig. \eqref{figMicromegas}. Nevertheless, imposing the perturbative limit $\lambda_\phi \lesssim 1$, the heavier fermion can only play a role in setting the DM abundance for $m_{\tilde \chi} \lesssim$ 10 TeV. 

  Models of DM with a light scalar mediator falls into the category of the so-called Self Interacting Dark Matter (SIDM) models, which in principle may alliviate some of the tensions of the standard cold DM paradigm when comparing numerical simulations with observations \cite{Tulin2013,Tulin2013a,Kaplinghat2013,Kaplinghat:2015aga,Kahlhoefer2017} (for a review see for example \cite{Tulin:2017ara}). The impact of DM self-interactions on astrophysical objects is quantified by the momentum transfer cross section $\sigma_T$ \cite{Kahlhoefer2017},
  \be
  \sigma_T= 2 \pi \int_{-1}^1 \frac{d \sigma}{d \Omega} ( 1 - | \cos \theta|) d \cos \theta \,,
  \ee
  where $\sigma$ is the DM scattering cross-section and $\theta$ the scattering angle in the center of mass frame. Viable SIDM models preferred a scattering cross section with a mild relative velocity dependence $v$ from dwarf to cluster of galaxies scales. For example to be consistent with data from the Bullet Cluster one needs $\sigma_T/m_{\chi} \lesssim 0.7$ cm$^2$/g at $v\simeq 4000$ km/s \cite{Randall:2008ppe}; similarly from other cluster observations with velocities $v\simeq O(1000-4000)$ km/s, one has $\sigma_T/m_\chi \lesssim {\rm few}\times 1$ cm$^2$/g \, \cite{Tulin:2017ara}. On the other hand, at smaller velocities $v\simeq O(10-500)$ km/s, typically one requires $\sigma_T/m_\chi \gtrsim 1$ cm$^2$/g. In our model, with $m_\chi \simeq 1$ GeV, $m_\phi \simeq 1$ MeV and $g_\chi \simeq 0.053$ we can compute analytically $\sigma_T$ within the Bohr approximation \cite{Kahlhoefer2017}, valid for $g_\chi^2 m_\chi/(4 \pi m_\phi) \ll 1$, and we have: $\sigma_T/m_\chi \simeq 0.032$ cm$^2$/g with $v=4000$ km/s; $\sigma_T/m_\chi \simeq 2.4$ cm$^2$/g with $v=1000$ km/s; $\sigma_T/m_\chi \simeq 36.2$ cm$^2$/g with $v=30$ km/s. For dwarf galaxies scales and $v \approx 30$ km/s, we seem to have slightly larger values than the one typically taken as a conservative upper limit $\sigma_T/m_\chi \simeq 10$ cm$^2$/g. However, given the uncertainties in the field, this upper limit might be too conservative, and a factor $O(3)$ larger could be acceptable. We also noticed that $\sigma_T \propto g_\chi^4$, and a factor $O(3)$ reduction in $\sigma_T$ only requires a coupling $g_\chi \simeq 0.04$. This value is still compatible with having the right DM abundance by standard freeze-out, by including the contribution from the heavier fermion with $m_{\tilde \chi} \leq 1$ TeV (see Fig. \eqref{figMicromegas}). Therefore, in the following sections we will fix $g_\chi \simeq 0.053$, and the parameter space available will be given for this value (in Fig. \eqref{plotlimitzchiznchi}), while in Appendix A we comment how different limits are changed when considering a smaller value $g_\chi \simeq 0.04$. 

\section{Scalar potential} \label{ScalarPotential}

We now check constraints on the parameters of the scalar potential from electroweak symmetry breaking and the limit on the invisible Higgs decays. This was done in Ref. \cite{Kouvaris2015}, and we follow here their approach for the parameter values we are interested in. 

The general potential between a singlet real scalar $S$ and a doublet $\Sigma$ is given by:
\bea
V(S, \Sigma)&=& m_\Sigma^2 | \Sigma|^2 + m_S^2 S^23 + \lambda |\Sigma|^4 + \frac{\lambda_S}{4} S^4 \nonumber \\
&&+ \frac{\lambda_{S\Sigma} }{2} S^2 | \Sigma^2| + \frac{\mu_3}{3} S^3 + \mu_1 S | \Sigma^2| \,,
\eea
where $m_\Sigma$, $m_S$ are mass parameters, $\mu_1$, $\mu_3$ trilinear couplings, and $\lambda$, $\lambda_S$, $\lambda_{S\Sigma}$ dimensionless quartic couplings. The physical states $h$ and $\phi$ are obtained through the mixing of the neutral component of the doublet $\sigma^0$ and the singlet $s$:
\bea
\sigma^0 &=& \cos \theta h + \sin \theta \phi \,, \\
s &=&  \cos \theta \phi - \sin \theta h \,,
\eea
with a mixing angle after electroweak symmetry breaking given by:
\be
\sin \theta \simeq \frac{v_{EW}}{m_h^2} ( \mu_1 + \lambda_{S\Sigma} w ) \,, \label{sintheta}
\ee
where $m_h=125$ GeV is the Higgs mass , $v_{EW}= 246.22$ GeV its vacuum expectation value (vev), and $w$ the singlet one. With $m_\chi = g_\chi w \simeq 1$ GeV, one has $w \lesssim 19$ GeV, i.e., $w \ll v_{EW}$. 

Given that $m_\phi \ll m_\chi \simeq 1$GeV,  the Higgs can now decay into light scalars and dark fermions with rates\footnote{We keep the notation $g_\chi$ for the coupling between the physical scalar $\phi$ and the fermion $\chi$, and therefore the coupling between the physical Higgs $h$ and $\chi$ is given by $g_\chi \tan \theta$.}:
\bea
\Gamma ( h \rightarrow \phi \phi) &=& \frac{\lambda_{S\Sigma}^2 v_{EW}^2}{8 \pi m_h} \left( 1 - \frac{4 m_\phi^2}{m_h^2}\right)^{1/2} \,, \label{gamhphi}\\
\Gamma ( h \rightarrow \chi \chi) &=& \frac{g_\chi^2 \tan \theta^2 m_h}{8 \pi} \left( 1 - \frac{4 m_\chi^2}{m_h^2}\right)^{3/2} \,. \label{gamhchi}
\eea
The invisible branching ratio must be less than 12\%  \cite{AtlasHiggs}, and using the SM Higgs decay rate $\Gamma_h= 4.1$ MeV, the decay into scalars sets an upper limit $\lambda_{S\Sigma} \lesssim 5 \times 10^{-3}$. Similarly from Eq. \eqref{gamhchi} and using $g_\chi \simeq 0.053$ we obtain the upper limit on the mixing angle $\tan \theta \lesssim 0.1982$  ($\sin \theta \lesssim 0.1944$). On the other hand, the decay rate of $\phi$ into a pair of $e^+e^-$ is given by:
\be
\Gamma_\phi = \frac{h_e^2\sin \theta^2}{8 \pi} m_\phi \left(1 - \frac{4 m_e^2}{m_\phi^2}\right)^{3/2} \,,
\ee
where $h_e =2.9 \times 10^{-6}$ is the SM electron Yukawa coupling. 
To be safe from BBN constraints \cite{Pitrou:2018cgg, Planck2020}, we need $\tau_\phi \lesssim 1$ s, which  only requires a mixing angle $\sin \theta \gtrsim 10^{-6}$ \cite{ArkaniHamed2009,Lin2012,Tulin2013,Kaplinghat2013,Kouvaris2015}.

Given that $w< v_{EW}$ and $\lambda_{S\Sigma} \ll 1$, the singlet practically does not contribute to the Higgs mass, $m_h^2 \simeq 2 \lambda v_{EW}^2$, and $\lambda \simeq  \lambda_{SM} \simeq 0.13$. 
For the trilinear coupling $\mu$ among physical states we have:
\be
\mu
\simeq 4 ( \mu_1 + \lambda_{S\Sigma} w ) \simeq 4 \frac{m_h^2}{v_{EW}} \sin \theta \,, \label{muapprox}
\ee
where we have used Eq. \eqref{sintheta}. 
For the repulsive interaction among neutrons and DM fermions we also need $\mu <0$, i.e. $\mu_1 < -\lambda_{S\Sigma} w$.
Therefore, from the limits on the mixing angle, the  absolute value of the trilinear coupling must lie in the range:
\be
2.5 \times 10^{-4} ~ \gev\lesssim | \mu| \lesssim 49.4~\gev \,. \label{limitmu}
\ee

\begin{table}[bp]
\begin{center}
	\begin{tabular}{|c|c|c|c|c|c|}
		\hline
		Model & Label & $a$ (MeV) & $\alpha$ & $b$ (MeV) & $\beta$ \\
		\hline
		none & QMC1 & 12.7 & 0.49 & 1.78 & 2.26 \\
		$V_{2\pi}^{PW} + V_{\mu=150}^R$ & QMC2 & 12.7 & 0.48 & 3.45 & 2.12 \\
		$V_{2\pi}^{PW} + V_{\mu=300}^R$ & QMC3 & 12.8 & 0.488 & 3.19 & 2.20 \\
		$V_{3\pi} + V_R$ & QMC4 & 13.0 & 0.49 & 3.21 & 2.47 \\
		$V_{2\pi}^{PW} + V_{\mu=150}^R$ & QMC5 & 12.6 & 0.475 & 5.16 & 2.12 \\
		$V_{3\pi} + V_R$ & QMC6 & 13.0 & 0.50 & 4.71 & 2.49 \\
		UIX & UIX & 13.4 & 0.514 & 5.62 & 2.436 \\ \hline
	\end{tabular}
	\caption{\small EoS given by the QMC predictions of two-body and three-body interactions. The second column shows the name used in this work to identify the different equations.}
	\label{tableeos}
\end{center}
\end{table}

\section{Neutron Stars and DM}\label{NSDM}

For a stellar compact object like a NS, the mass-radius relation can be obtained by integrating Einstein equations for a static, spherically symmetric configuration, the Tolman-Oppenheimer-Volkoff (TOV) equations \cite{tov1,tov2}.
\bea
\frac{d P(r)}{dr}\!\!&=&\!\! - \frac{G (\rho(r)+ P(r))}{r^2}\cdot \frac{M(r) + 4 \pi P(r) r^3}{1 - 2 G M(r)/r} \,, \label{dPr}\\
\frac{d M(r)}{dr}\!\!&=& \!\!4 \pi r^2 \rho(r) \label{dMr}\,,
\eea
where $G$ is the Newton constant, $P$ the pressure, $\rho$ the energy density, $M$ the stellar mass and $r$ the radius. TOV Eqs. are integrated for a given initial pressure (or number density) at the center of the object, and a given equation of state, i.e., the relation between $P$ and $\rho$, until we reach the surface pressure $P(R)=0$, at which point we also get  the total mass and the maximal radius $R$.

Astrophysical observations in principle can be used to put constraints on the EoS of the NS, excluding those that are not able to reach the $2 M_\odot$ limit, the largest observed NS mass \cite{MassNS}. 
Adding the new neutron decay channel, one has to take into account the conversion of neutrons into DM fermions inside the NS. And typically, the addition of the DM particle effectively modifies the EoS inside the stellar object such that the maximum NS mass falls well below the observational limit. One possible solution is to allow for repulsive DM interactions \cite{Cline2018, Grinstein2019}, which disfavours the conversion. 

There is a variety of EoS for nuclear matter in the literature \cite{Pieper:2001ap,Wiringa:2002ja,Gandolfi:2011xu,Hebeler:2013nza}, and 
in order to check how the DM hypothesis modifies the NS mass-radius relation we have chosen to work with a simple parametrization of the internal energy per neutron:
\be
E_n(x) = a x^\alpha + b x^\beta \,,
\ee
where $x=n_n/n_0$, $n_n$ the nucleon density and $n_0=0.16 ~{fm}^{-3}$ the saturation density. Different sets of  parameters correspond to different choices of the EoS, and we work with those given in Table \ref{tableeos}. 
Energy density and pressure are given by $\rho_n= n_n(m_n + E_n)$ and $P_n= n_n^2 d E_n/d n_n$ respectively.
  
\begin{table}[b]
\centering
\begin{tabular}{| c | c c c |} \hline
	Object & Mass ($M_{\odot}$) & Radius (km) & \\ \hline
	\rule{0pt}{12pt}
	Pulsar PSR J0030+0451 & $1.44^{+0.15}_{-0.14}$ & $13.02^{+1.24}_{-1.06}$ & \cite{Miller2019} \\ \hline
	\rule{0pt}{12pt}
	\multirow{2}{*}{Pulsar PSR J0740+6620} & $2.07^{+0.07}_{-0.07}$ & $12.39^{+1.30}_{-0.98}$ & \cite{Riley2021} \\
	\rule{0pt}{12pt}
	& $2.08^{+0.09}_{-0.09}$ & $13.71^{+2.61}_{-1.50}$ & \cite{Miller2021} \\ \hline
	\rule{0pt}{12pt}
	\multirow{2}{*}{GW190814 \cite{Abbott2020}} & $2.59^{+0.08}_{-0.09}$ & $14.77 - 14.87$ & \cite{KanakisPegios2021} \\
	\rule{0pt}{12pt}
	& & $11.35 - 13.67$ & \cite{Miao2021} \\ \hline
\end{tabular}
\caption{\small NS mass and radius observations with NICER (pulsars) and LIGO (GWs). It is not completely confirmed if the compact object in event GW190814 is the heaviest NS or a small black hole.}
\label{tableNicer}
\end{table}

\begin{figure}[t]
\centering
\includegraphics[width=8.5cm]{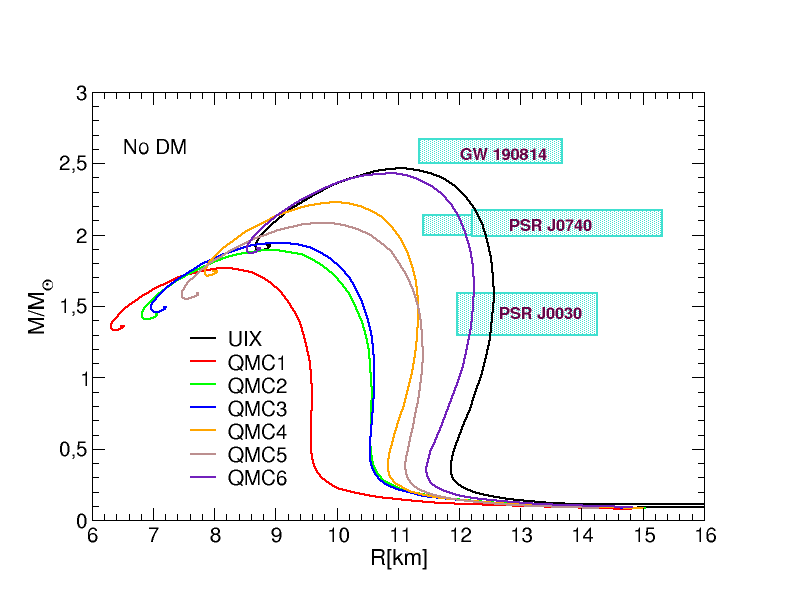}
\includegraphics[width=8.5cm]{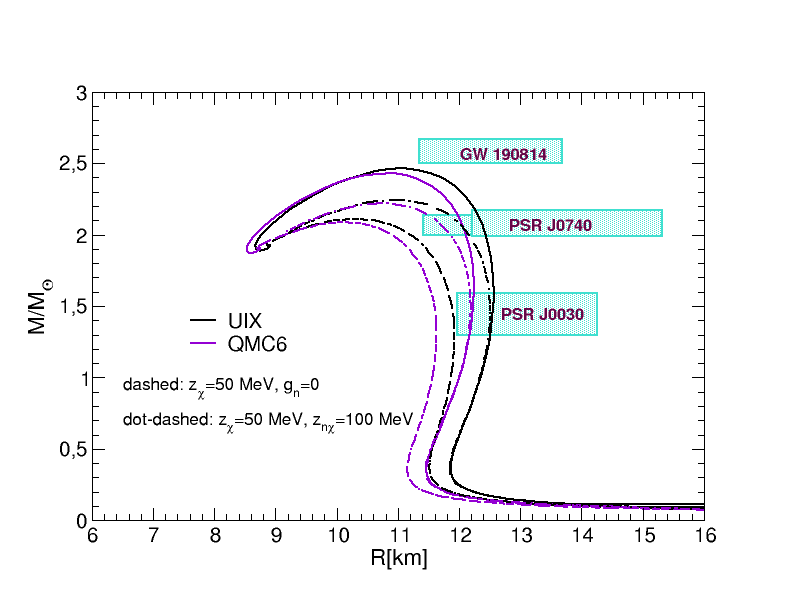}
\caption{\small Top: Mass-radius relation of NS described by the family of EoS given in table \ref{tableeos}, without DM. Although the differences among parameters of the same type are rather small, the resulting values for each EoS have non-negligible changes. The coloured contours represent the observational values on masses and radii from NICER and LIGO.
  Bottom: Mass-radius relation including DM repulsive and self-interactions, as indicated by the value of $z_{n\chi}$ and $z_\chi$ in the plot; $g_n=0$ means that no repulsive interaction is included. We only considered as examples the EoS QMC6 and UIX.}
\label{plotMRDM}
\end{figure}

\begin{figure*}[t]
\centering
\includegraphics[width=12cm]{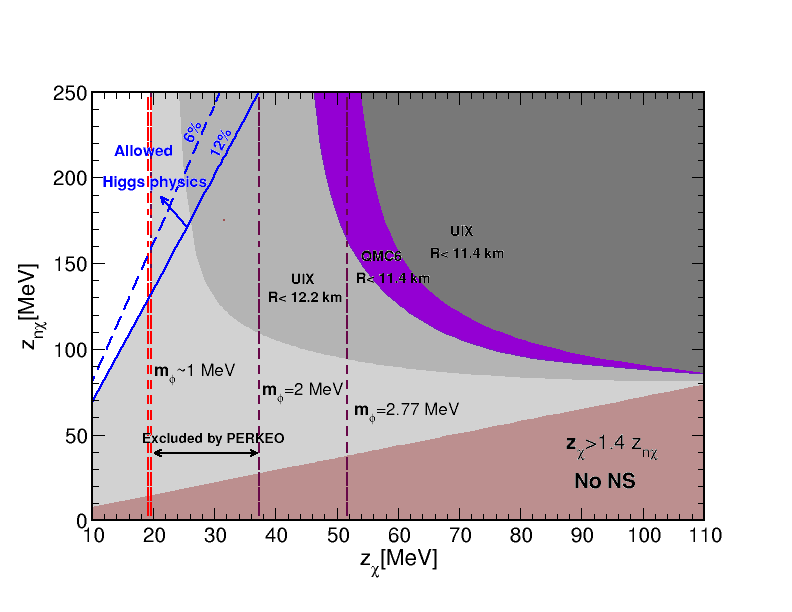}
\caption{\small Parameter space available in the plane $z_\chi-z_{n\chi}$ when including DM. Grey areas labeled ``UIX'' are excluded when using the UIX EoS because they do not give rise to 2 solar masses NS with a large enough radius with  
either $R > 11.4$ km (dark grey) or $R > 12.2$ km (light grey). Similarly for the violet shaded region for the QMC6 EoS. The brown shaded region on the bottom is excluded because the large repulsive interaction will give rise to a ``DM'' star instead of a NS. This bound is independent of the particular EoS used for nuclear matter. Vertical solid lines are the values of $z_\chi$ for a given value of $m_\phi$ as indicated in the plot, and $g_\chi\simeq 0.053$ as required to have the right DM abundance obtained in section \ref{DMabundance}. On the left, the narrow band between vertical red dashed lines labeled $m_\phi \sim 1$ MeV is the range {\it not} excluded by PERKEOII experiment. The left hand corner solid (dashed) blue line is the lower limit on $z_{n\chi}$ from Higgs physics, with Br(Higgs $\rightarrow$ invisible)=12 \% (6 \%). The white region above this line is the parameter space available fulfilling all constraints considered. See the text for details. 
}
\label{plotlimitzchiznchi}
\end{figure*}

We have first integrated the TOV Eqs. \eqref{dPr}-\eqref{dMr} without DM, for the set of EoS given in Table \eqref{tableeos}. The mass-radius curves are plotted in Fig. \eqref{plotMRDM} (top panel). We have also included observational limits from NICER collab. \cite{Miller2019,Riley2019, Miller2021, Riley2021}, and the not yet confirmed limit from LIGO collab. GW190814 \cite{Abbott2020}. These examples serve us as a guide to understand the physics when including DM. Only the EoS UIX and QMC6 are able to reproduce NS with  $2M_\odot$ and a radius within the observable range, and we focus therefore on these cases, adding DM with self-interactions and repulsive $n-\chi$ interactions. The total energy density of the NS is now given by:
\be
\rho_T( n_n, n_\chi) = \rho_n(n_n) + \rho_\chi^{(0)}(n_\chi) +  \frac{n_\chi^2}{2 z_\chi^2} +\frac{ n_n n_\chi}{2 z_{n \chi}^2} \,, \label{rhoT}
\ee
where $\rho_n$ is the neutron energy density, $n_\chi$ the dark fermion number density, the third term  is the contribution due to DM self-interactions, and the contribution from the repulsive potential is given by the last term;
$\rho_\chi^{(0)}$ is the energy density of the DM fermion without self-interactions, modelled as that of a free Fermi gas, 
\be
\rho_\chi^{(0)}=\frac{m_\chi^4}{8 \pi^2} \left( x \sqrt{1 + x^2} ( 1 + 2 x^2) - \ln ( x + \sqrt{1+ x^2}) \right) \,, \label{rhox} 
\ee
with $x=k_F/m_\chi=(3 \pi^2 n_\chi)^{1/3}/m_\chi$, $k_F$ being the Fermi momentum.
In term of masses and couplings at the particle physics level given in Eq. \eqref{Lparticles}, the variables $z_i$ are given by: 
\be
z_{n\chi}= \frac{m_\phi}{\sqrt{g_\chi |g_n|}}\,, \;\;\;
z_\chi= \frac{m_\phi}{g_\chi}\,.
\ee
When integrating TOVs Eqs with DM self-interactions and repulsive interactions, the value of the Fermi momentum (DM number density) is derived from the chemical equilibrium condition:
\be
\sqrt{k_F^2 + m_\chi^2} + \left(\frac{1}{z_\chi^2} - \frac{1}{2z_{n\chi}^2}\right) n_\chi = \frac{\partial \rho_n}{\partial n_n} - \frac{n_n}{2z_{n\chi}^2}  \,. \label{ChemEq}
\ee  

The presence of DM, even with self-interactions, tends to soften the effective EoS of the neutron star, given rise to smaller and less massive NS \cite{Cline2018}. This effect can be compensated by the effective repulsive interaction parametrised by $z_{n\chi}$. This can be seen in the example plotted in the bottom panel in Fig. \eqref{plotMRDM}, for the EoS UIX and QMC6. We include as solid lines the case without DM for comparison, one without repulsive interactions ($g_n=0$, dashed lines), and including the latter with $z_{n \chi}= 2 z_\chi= 100 $ MeV (dot-dashed lines).  For values of the couplings such that $z_{n\chi} \geq z_\chi/\sqrt{2}$, the chemical equilibrium condition favours configurations with $n_\chi \ll n_n$ (or no DM at all). However, when $z_{n\chi} < z_\chi/\sqrt{2}$ and the second term in Eq. \eqref{ChemEq} changes sign, a configuration with $n_\chi \gg n_n$, or practically no neutrons at all, is preferred, i.e., something more like a ``DM'' star. Therefore, independently of the EoS of nuclear matter, this imposes a lower bound on  $z_{n\chi} \gtrsim z_\chi/\sqrt{2}$ in order to have a stellar compact object with neutrons as the main component. 

  Varying $z_\chi$, $z_{n_\chi}$ we then obtain the parameter space available in order to reproduce NS with 2 solar masses and a radius larger than 11.4 km \cite{Riley2021} (12.2 km \cite{Miller2021}), shown in Fig. \eqref{plotlimitzchiznchi}. We stress that although the upper limit on the repulsive parameter $z_{n\chi}$ depends on the EoS used for nuclear matter, and may disappear or be relaxed for other parametrizations of the EoS, the lower limit is universal. We also notice that the larger the value of the self-interacting parameter $z_\chi$ the more difficult to get massive enough neutron stars due to the softening of the EoS, and although this could be compensated by increasing the repulsive parameter $z_{n\chi}$, this is not possible beyond some value $z_\chi \approx 110$ MeV because the repulsive interaction turns the star into a DM one. Therefore, once the value of $g_\chi$ is set in order to obtain the right DM abundance, this would always translate into an upper bound for the light scalar mediator mass. Using the value of the coupling $g_\chi \simeq 0.053$ obtained in section \ref{DMabundance}, we can set the value of $z_\chi$ for different $m_\phi$ masses. Those are indicated by the vertical dashed lines in the figure. The narrow vertical band between the first two red dashed lines to the left corresponds to the range $2 m_e < m_\phi < 2 m_e + 0.0375 \, \mev$, not excluded yet by the experiments \cite{Klopf:2019afh}. Finally, the solid blue line at the left upper corner indicates the lower limit on the parameter $z_{n\chi}$ set by the upper limit on the trilinear coupling $\mu$, obtained in section \ref{ScalarPotential} from the current limit on the branching ratio of the invisible Higgs decay, such that:
\be
126.4 \lesssim \frac{z_{n \chi}}{m_\phi} \lesssim 5.6 \times 10^4 \,. \label{znchirange} 
\ee
For comparison, we also include with a dashed blue line how this limit would be affected with a more restrictive limit say Br(Higgs $\rightarrow$ invisible)=6 \%.
This would reduce the maximum allowed value for the singlet-Higgs mixing angle, and then the trilinear parameter $\mu$, which increases the slope of the limit in the $z_{n\chi}-z_\chi$ plane. Combining all constraints, we are left with the values of the parameters in the white area. In the plot we have chosen not to exclude values of $m_\phi < 2 m_e$, but we remark that to access this region we need to extend the model such that the light scalar decays before BBN.  

The allowed range given in Eq. \eqref{znchirange} can be expressed as limits on the neutron-light mediator coupling $g_n$:
\be
6\times 10^{-9} \lesssim |g_n| \lesssim 1.2 \times 10^{-3} \,. 
\ee
This coupling is constrained by supernova cooling bounds \cite{Dent:2012mx,Mahoney:2017jqk,Calore:2021lih,Heeck:2014zfa,Dev:2020jkh,Balaji:2022noj} when the light mediator mass falls in the range $[1,\, 100]$ MeV. The coupling must be small enough in order to avoid excesive bremsstrahlung production of the light scalar, which  can lead to a rapid cooling of the progenitor core. On the other hand, production and absorption processes compensate each other when increasing the coupling, with no net cooling effect beyond some value.  Given the bound obtained in \cite{Balaji:2022noj} on the singlet-Higgs mixing angle, which excludes the range $10^{-7} \lesssim \sin \theta \lesssim 3 \times 10^{-5}$, it would be sufficient to have $|g_n| \gtrsim 1.8 \times 10^{-7}$ to avoid these limits and then  $z_{n\chi} \lesssim 10^{4}$, which only modifies the upper limit in Eq. \eqref{znchirange} by approximately a factor of $O(6)$.

\vspace{1.3cm}

\section{Summary and discussion} \label{Summary}
Let us summarize the values of masses and couplings for the model considered, from different physical considerations:
\begin{itemize}
\item {\bf Neutron decay anomaly:} The neutron decay anomaly could be explained with a new neutron decay channel into a dark fermion and scalar. An explicit model was given in \cite{Grinstein2019}, Eq. \eqref{Lparticles}. Masses must be in the range given in Eqs. \eqref{masses1} and \eqref{masses2}, to allow neutron decay but ensuring the stability of nuclei and the proton. We have chosen to work with a dark fermion mass $m_\chi \simeq 1~ \gev$, and a light scalar $m_\phi \sim O(\mev)$, but larger than $2 m_e$. This allows the scalar to decay into a pair electron-positron, with a small enough lifetime to avoid problems at BBN. This will reintroduce the neutron decay channel $n \rightarrow \chi + e^- + e^+$, excluded so far only for certain values of the kinetic energy $E_{e^-e^+}$. We can then adjust the value of $m_\phi$ in order to avoid those limits. Other particles in the model, like $\tilde \chi$ and $\Phi$ are assumed to be heavy enough, say O(1 TeV). 

\item {\bf Dark matter abundance:} We have checked semi-analytically and with micrOMEGAS that a dark fermion with a mass $m_\chi \simeq 1$ GeV is a good candidate for dark matter, and reproduces the present dark matter abundance with a coupling value $g_\chi\simeq 0.053$, and $\lambda_\phi/\sqrt{m_{\tilde \chi}} \ll 0.01$  GeV$^{-1/2}$. 

\item{\bf Higgs physics:} With the inclusion of a light singlet scalar, the standard model Higgs can now decay into both light scalars or dark fermions. Demanding that the invisible branching ratio be less than 12\% sets an upper bound on the singlet-Higgs mixing angle, while a lower bound can be  set to ensure
the decay of the light scalar before BBN: $10^{-6} \lesssim |\sin \theta| \lesssim 0.1944$. This translates into the allowed range for the trilinear interaction given in Eq. \eqref{limitmu}. However, supernova cooling bounds give a larger lower limit on the mixing angle, $\sin \theta \gtrsim 3\times 10^{-5}$, and then
$ 7.5 \times 10^{-3} ~ \gev\lesssim | \mu| \lesssim 49.4~\gev \,$.

\item {\bf Neutron Stars:} The model in \cite{Grinstein2019} gives rise to  self-interactions for the dark fermion, through the exchange of the light scalar, and an effective repulsive interaction neutron-dark fermion through the Higgs portal. Both interactions will affect the NS mass-radius relation, derived by integrating the TOV Eqs. Having fixed $m_\chi$, $m_\phi$ and $g_\chi$, the parameter for the self-interactions is given by $z_\chi \simeq 19 m_\phi ~\mev$, which implies a lower limit for the repulsive parameter $z_{n \chi}^{min} \gtrsim 13.3 m_\phi~ \mev$, independently of the EoS. However, Higgs physics constrains the range of possible trilinear singlet-Higgs interactions, and therefore sets a more restrictive lower limit $z_{n \chi} \gtrsim 126.4 m_\phi $.
\end{itemize}

All constraints together, for the EoS we have considered, demanding to have NS with $M_{NS} \simeq 2 M_{\odot}$ and $R > 12.2$ km \cite{Miller2021} just  leaves a very narrow range of possible values for the light scalar mass $2 m_e < m_\phi < 2 m_e + 0.0375 \, \mev$ (see Fig. \eqref{plotlimitzchiznchi}). On the other hand, only requiring to have massive enough NS with $R > 11.4$ km, would open the parameter space for the UIX (QMC6) EoS for slightly larger masses $m_\phi \simeq 2-2.77$ MeV. In any case, NS physics alone sets an upper limit on the light scalar $m_\phi \lesssim 6$ MeV, which gets further reduced when combined with the constraints on the singlet-Higgs mixing. 

The lower limit on $m_\phi$ comes from the minimal requirement to ensure the decay of $\phi$ before BBN into a pair $e^+e^-$. Smaller values for $m_\phi$ might be possible but this requires extending the model as in the original proposal \cite{Fornal2018}. Nevertheless, the model with the singlet decaying into SM particles is still viable. 

\section*{Acknowledgments}
The work of MBG has been supported by grant PID2022-140831NB-I00 funded by MICIU/AEI/10.13039/501100011033 and FEDER, UE, as well as FCT - Fundacao para a Ciencia e Tecnologia, I.P. through the project CERN/FIS-PAR/0027/2021, with DOI identifier 10.54499/CERN/FIS-PAR/0027/2021. THR acknowledges support from grant PID2020-115758GB-I00/PRE2021-100042 financed by MCIN/AEI/10.13039/501100011033 and the European Social Fund Plus (ESF+). The authors acknowledge funding from the French Programme d’investissements d’avenir through the Enigmass Labex.

\begin{figure*}[t]
\centering
\includegraphics[width=12cm]{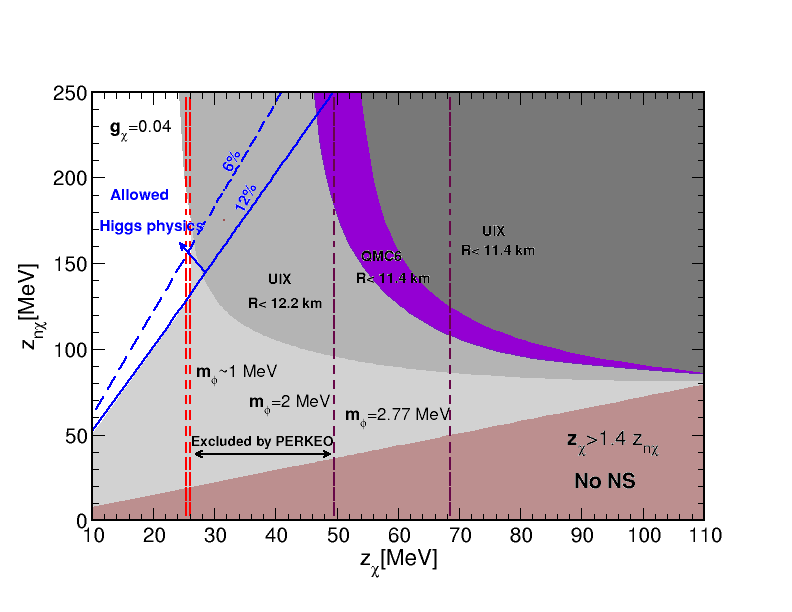}
\caption{\small Parameter space available in the plane $z_\chi-z_{n\chi}$ when including DM, same than Fig. \eqref{plotlimitzchiznchi} but for $g_\chi\simeq 0.04$. Vertical lines for fixed values of $z_\chi$ are now slightly displaced towards the right, and the slope of lower limit on $z_{n\chi}$ from Higgs physics is smaller. As before, the white region is the parameter space available.}
\label{plotlimitzchiznchigx004}
\end{figure*}

\appendix

\section{DM abundance and SIDM constraints: changes in parameter limits}
As mentioned at the end of section \ref{DMabundance}, a value $g_\chi \simeq 0.053$, which gives today's dark matter abundance when $\lambda_\phi/\sqrt{m_{\tilde \chi}} \ll 0.01$ GeV$^{-1/2}$, seems in conflict with the conservative bound on the average scattering cross-section at dwarf galaxies scales, $\sigma/m_\chi \lesssim 10$ cm$^2$/g, and slightly smaller values might be preferred; for example  $g_\chi \simeq 0.04$, which is also consistent with the DM abundance when $\lambda_\phi/\sqrt{m_{\tilde \chi}} \simeq 0.033$ GeV$^{-1/2}$ and $m_{\tilde \chi} \lesssim 1$ TeV. Changing the value of $g_\chi$ affects different parameter values, and finally the available parameter space when combining all limits. In Table \ref{tablelimits} we compare different upper limits: the values of $\tan \theta$ and $\mu$ goes like the inverse of $g_\chi$, while $|g_n| \propto 1/\sqrt{g_\chi}$. The lower limit in Eq. \eqref{znchirange} also changes to $z_{n \chi}/m_\phi \gtrsim 127.9 $. And in Fig. \eqref{plotlimitzchiznchigx004} we show the available parameter space (white area)  for $g_\chi \simeq 0.04$. 

For our preferred light scalar mass range $2 m_e < m_\phi < 2 m_e + 0.0375 \, \mev$, the value $g_\chi \simeq 0.04$ is almost in conflict with NS physics for the EoS we have considered in this work. Indeed, it would be interesting to work with a more realistic NS EoS to check first whether this could sets a lower limit on $g_\chi$. Having set this coupling value, the next step would be to compare with data from small scale structure, beyond conservative limits, which we leave for a future work.

\begin{table}[htbp]
\begin{center}
	\begin{tabular}{|c|c|c|}
		\hline
		Upper limit & $g_\chi=0.053$ & $g_\chi=0.04$ \\
		\hline
		$\tan \theta$ & 0.1982 & 0.2626 \\
                $\sin \theta$ & 0.1944 & 0.254 \\
		$\mu$ & 49.4 GeV & 64.5 GeV \\
                $|g_n|$ & $1.2\times 10^{-3}$ & $1.5\times 10^{-3}$  \\
\hline                
	\end{tabular}
	\caption{\small Upper limits on different parameter values depending on the value of $g_\chi$.}
	\label{tablelimits}
\end{center}
\end{table}


\section{Neutron decay}
For completeness, we summarize here the expressions of the neutron decay rate for the different dark channels.

\hspace{0.1cm}

 (a) {\bf Decay into scalar and fermion}: Starting from the lagrangian in Eq. \eqref{Lparticles},  and integrating out the heavy states, neutron effective interactions with a dark scalar $\phi$ and fermion $\chi$  are given by the effective lagrangian
\be
\calL_{n\rightarrow \chi\phi}^{eff}= \frac{\lambda_\phi \epsilon}{ m_n - m_{\tilde \chi}} \bar \chi ~n~ \phi^* \,,
\ee
with $\epsilon=\frac{\beta \lambda_q \lambda_\chi}{M_\Phi^2}$ and $\beta=0.0144(3)~\gev^3$ from lattice calculations. The decay rate is then; 
\be
\Gamma_{n\rightarrow \chi\phi} = \frac{ \lambda_\phi^2 \epsilon^2}{16 \pi} \sqrt{f(x,y)} \frac{m_n}{(m_n-m_{\tilde \chi})^2}\,,
\ee
where $x=m_\chi/m_n$, $y=m_\phi/m_n$, and:
\be
\sqrt{f(x,y)} \!\!= \!\!( (1-x)^2 - y^2)^{1/2} ((1+x)^2 -y^2)^{3/2} \,.
\ee
 Working in the limit $m_{\tilde \chi} \gg m_n$, we can write:
\be
\Gamma_{n\rightarrow \chi\phi} = \frac{\kappa_\phi^2 }{16 \pi} \sqrt{f(x,y)} m_n \,, \label{gammanphi}
\ee
with
\be
\kappa_\phi \simeq \lambda_\phi \lambda_q \lambda_\chi\cdot \frac{\beta}{ m_{\tilde \chi} M_\Phi^2} \simeq 1.44 \times 10^{-11} \left(\frac{\tev^3}{m_{\tilde \chi} M_\Phi^2} \right)\lambda_\phi \lambda_q \lambda_\chi \,. 
\ee
For example for $m_\phi= 2 me + 0.0375$ MeV and $m_\chi= 938.513~ \mev$ (upper limit from Eq. \eqref{masses1}), we have: 
\be
Br(n\rightarrow \chi\phi) \simeq 16.3 ~ (\lambda_\phi \lambda_q \lambda_\chi)^2 \left (\frac{\tev^3}{ m_{\tilde \chi} M_\Phi^2}\right) \,,
\ee
and with $m_\chi= 936.941~ \mev$ (lower limit): 
\be
Br(n\rightarrow \chi\phi) \simeq 10^{8}~ (\lambda_\phi \lambda_q \lambda_\chi)^2 \left( \frac{\tev^3}{ m_{\tilde \chi} M_\Phi^2}\right) \,.
\ee
Therefore, having $Br(n \rightarrow \chi\phi) \simeq 0.01$ would only require couplings $O(0.01-0.2)$.

\hspace{0.1cm}



\hspace{0.1cm}

(b) {\bf Decay into dark fermion and electron-positron mediated by a light scalar}: $ n \rightarrow \chi + \phi \rightarrow \chi + e^- + e^+$

When $m_\phi \gtrsim 1~\mev$, we re-introduce the decay channel into a pair electron-positron. The 3-body decay rate is given by:
\be
\frac{d \Gamma_3}{d E_\chi} = \frac{1}{64 \pi^3 m_n} | {\cal M}|^2 \cdot \frac{1}{q^2} (E_\chi^2 - m_\chi^2)^{1/2} ( q^2 ( q^2 - 4 m_e)^2)^{1/2} \,,
\ee
where $q^2= m_n^2 - 2 E_\chi m_n + m_\chi^2$ is the momentum transferred, and ${\cal M}$ the scattering amplitude:
\begin{widetext}
\be
| {\cal M} |^2 = 4 \bar \kappa_e^2 m_n (  E_\chi + m_\chi) ( m_n^2 + m_\chi^2 -2 m_n E_\chi - 4 m_e^2) \cdot \frac{1}{ (q^2 - m_\phi^2)^2 + m_\phi^2 \Gamma_\phi^2} \,.
\ee
\end{widetext}
We have defined the coupling $\bar \kappa_e^2= \kappa_\phi^2 h_\phi^2$, and included the Breit-Wigner propagator for the light scalar, with decay rate:
\be
\Gamma_\phi= \frac{h_\phi^2}{8 \pi}\cdot m_\phi \cdot \left(1 - \frac{4 m_e^2}{m_\phi^2}\right)^{3/2} \,. \label{gammaphi}
\ee
Using the narrow width approximation,
\be
\frac{1}{ (q^2 - m_\phi^2)^2 + m_\phi^2 \Gamma_\phi^2} \rightarrow \frac{\pi}{m _\phi \Gamma_\phi} \delta( q^2 - m_\phi^2) \,,
\ee
we obtain:
\be
\Gamma_3 = \frac{\bar \kappa_e^2}{16 \pi^2 \Gamma_\phi} 
\cdot \frac{m_\phi}{2 m_n} \cdot (1 - \frac{4 m_e^2}{m_\phi^2})^{3/2} \cdot ( \tilde E_\chi - m_\chi)^{1/2} ( \tilde E_\chi + m_\chi)^{3/2} \,,
\ee
where we have defined: $\tilde E_\chi = (m_n^2 + m_\chi^2 - m_\phi^2)/(2m_n)$.
And replacing the scalar decay rate Eq. \eqref{gammaphi} we have:
\be
\Gamma_3 = \frac{\kappa_\phi^2}{16 \pi} \cdot m_n \cdot \left(\frac{ ((m_n - m_\chi)^2 - m_\phi^2)^{1/2} ((m_n + m_\chi)^2 - m_\phi^2)^{3/2}}{m_n^2} \right) \,,
\ee
which is equal to the decay rate $\Gamma_{n \rightarrow \chi \phi}$ Eq. \eqref{gammanphi}. If we want the model to explain the neutron decay anomaly, the branching ratio has to be O(1\%), which requires either $m_\phi \lesssim 2 m_e + 0.0375$ MeV, or larger masses $m_\phi \gtrsim 2 m_e +0.644 $ MeV. 

\vspace{2cm}

\bibliography{biblioNSDM}

\end{document}